\documentclass[twocolumn,footinbib,floatfix,aps,10pt,pra,longbibliography,superscriptaddress]{revtex4-1}
\usepackage[usenames,dvipsnames]{xcolor} 
\usepackage[normalem]{ulem}

\usepackage[T1]{fontenc}
\usepackage{amsmath,amssymb}
\usepackage{amsfonts}
\usepackage{bm}
\usepackage{bbm}
\usepackage{color}
\usepackage{latexsym}
\usepackage[english]{babel}
\usepackage{stmaryrd}
\usepackage{psfrag,graphicx}
\usepackage{subfigure}
\usepackage{natbib}
\usepackage{appendix}
\usepackage{enumitem}

\makeatletter
\def\@bibdataout@aps{%
 \immediate\write\@bibdataout{%
  @CONTROL{%
   apsrev41Control%
   \longbibliography@sw{%
    ,author="08",editor="1",pages="1",title="0",year="0"%
   }{%
    ,author="08",editor="1",pages="0",title="",year="1"%
   }%
  }%
 }%
 \if@filesw
  \immediate\write\@auxout{\string\citation{apsrev41Control}}%
 \fi
}%
\makeatother

\definecolor{mygrey}{gray}{0.35}
\definecolor{myblue}{rgb}{0.2,0.2,0.8}
\definecolor{myzard}{cmyk}{0,0,0.05,0}
\definecolor{mywhite}{rgb}{1,1,1}
\definecolor{myred}{rgb}{0.9,0.1,0.}
\usepackage[colorlinks=true,citecolor=myblue,linkcolor=myblue,urlcolor=myblue]{hyperref}

\newcommand{\tr}{\operatorname{{tr}}} 
\newcommand{\ket}[1]{\vert #1 \rangle} 
\newcommand{\bra}[1]{\langle #1 \vert} 
\newcommand{\id}{\mathbbm{1}}

\newcommand{\e}{\mathrm{e}}
\newcommand{\ii}{\mathrm{i}}

\begin{document}


\title{Scalable reconstruction of unitary processes and Hamiltonians}

\newcommand{\uniulmitp}{Institut f{\"u}r Theoretische Physik,
  Albert-Einstein-Allee 11, Universit{\"a}t Ulm, 89069 Ulm, Germany}

\author{M. Holz{\"a}pfel}
\affiliation{\uniulmitp}
\author{T. Baumgratz}
\affiliation{\uniulmitp}
\affiliation{Clarendon Laboratory, Department of Physics,
  University of Oxford, OX1 3PU Oxford, United Kingdom}
\author{M. Cramer}
\affiliation{\uniulmitp}
\author{M.B. Plenio}
\affiliation{\uniulmitp}


\begin{abstract}
  Based on recently introduced efficient quantum state tomography
  schemes, we propose a scalable method for the tomography of unitary
  processes and the reconstruction of one-dimensional local
  Hamiltonians.  As opposed to the exponential scaling with the number
  of subsystems of standard quantum process tomography, the method
  relies only on measurements of linearly many local observables and
  either (a) the ability to prepare eigenstates of locally
  informationally complete operators or (b) access to an ancilla of
  the same size as the to-be-characterized system and the ability to
  prepare a maximally entangled state on the combined system. As such,
  the method requires at most linearly many states to be prepared and
  linearly many observables to be measured. The quality of the
  reconstruction can be quantified with the same experimental
  resources that are required to obtain the reconstruction in the
  first place.  Our numerical simulations of several quantum circuits
  and local Hamiltonians suggest a polynomial scaling of the total
  number of measurements and post-processing resources.
\end{abstract}

\maketitle

\section{Introduction}

Quantum process tomography
\cite{Nielsen2007,Chuang1997,Poyatos1997,Dariano2001,Dur2001,
  Altepeter2003} is {\it the} standard for the verification and
characterization of quantum operations on well-controlled quantum
systems. Among others, recent experimental demonstrations of quantum
simulators of multi-partite quantum systems
\cite{Jurcevic2014,Richerme2014} have demonstrated that, by now, the
number of well controllable qubits is in a regime for which
conventional tomography techniques fail as the required experimental
and numerical post-processing resources scale exponentially with the
number of qubits. While there has been a considerable effort to
introduce scalable techniques that allow for an efficient
reconstruction \cite{Cramer2010, Baumgratz2013, Baumgratz2013a,
  Landon-Cardinal2012, Toth2010, Schwemmer2014, Steffens2014,
  Steffens2014a, Banaszek2013} and verification \cite{Silva2011,
  Flammia2011, Kim2014} of quantum states, quantum process tomography
still leaves much to be desired.

\begin{figure}[t]
  \begin{center}
    \includegraphics[width=0.95\columnwidth]{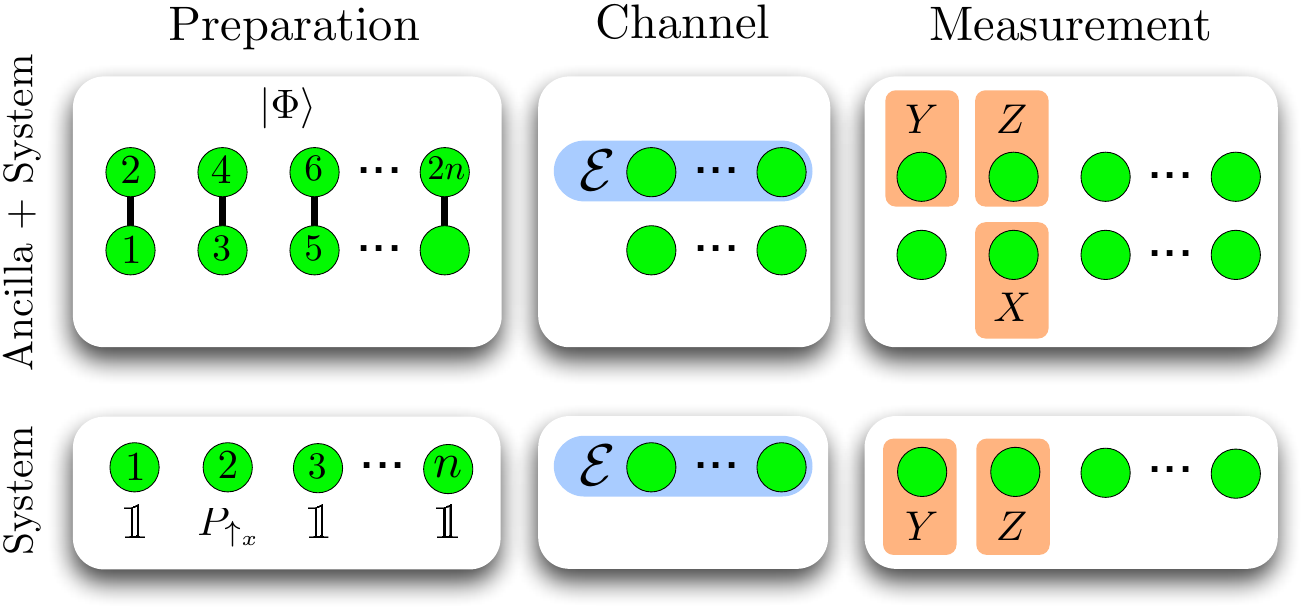}\vspace{-0.3cm}
  \end{center}
  \caption{(Color online) Two equivalent ways to obtain the necessary
    measurements for the scalable process tomography
    scheme. \textit{Top row:} System and ancilla are prepared in a
    maximally entangled state
    $|\Phi\rangle\propto\bigotimes_{i=1}^{N}[|0\rangle_{2i-1}|0\rangle_{2i}+|1\rangle_{2i-1}|1\rangle_{2i}]$
    (with pairwise entanglement indicated by vertical lines), the
    channel $\mathcal{E}$ is applied on the system, and products of
    Pauli observables are measured on all blocks of $r$ consecutive
    sites (exemplified for $r=3$ and
    $\hat{\sigma}_2^y\otimes\hat{\sigma}_3^x\otimes\hat{\sigma}_4^z$),
    which serves as input to reconstruct the resulting state
    $\hat{\varrho}_{\mathcal{E}}$ and thus the channel~$\mathcal{E}$.
    \textit{Bottom row:} Equivalently, these expectation values may be
    obtained without the ancillary system. The initial states have
    eigenstates of Pauli matrices (here, $\hat{P}_{\uparrow_x}$,
    $\hat{P}_{\downarrow_x}$, so those of $\hat{\sigma}_2^x$) on those
    sites where a Pauli matrix would have been measured on the
    ancilla. After application of the channel, the system part of the
    ancilla + system measurement is carried out (here, that of
    $\hat{\sigma}_1^y\otimes\hat{\sigma}_2^z$).}
    \label{fig:measurement-scheme}
\end{figure}

The most straightforward approach to process tomography is based on
the idea of probing the quantum channel with an informationally
complete set of states. After sending each of these input states
through the channel, the process underlying the dynamics is
characterized by performing full state tomography on all of the output
states. The so-obtained map fully characterizes the channel
\cite{Nielsen2007,Chuang1997,Poyatos1997}. This strategy is referred
to as \textit{standard quantum process tomography} (SQPT) and works
well as long as the underlying system is of low dimension. Considering
multi-partite quantum systems, however, reveals the disadvantages of
this technique: Not only the number of states to be sent through the
channel and to be indentified grows exponentially with the number of
subsystems $n$, but also the number of parameters that need to be
determined for each of the output states. There is, however, a
strategy to overcome the exponential scaling of the number of input
and output states: \textit{ancilla-assisted process tomography} (AAPT)
\cite{Dariano2001,Dur2001, Altepeter2003, ft:aapt-other-schemes}. AAPT
requires an ancillary system of the same size $n$ as the system
itself, the preparation of a maximally entangled input state
$\ket{\Phi} \propto \sum_{i}\ket{i}\ket{i}$ on the combined system,
and full tomography of the resulting state after application of the
channel,
$\hat{\varrho}_{\mathcal E} = (\id \otimes \mathcal
E)(\ket{\Phi}\!\bra{\Phi})$,
see Fig.\ \ref{fig:measurement-scheme}.  This method is based on the
correspondence between quantum states and quantum processes known as
the Choi-Jamio{\l}kowski isomorphism
\cite{Choi1975,Jamiolkowski1972}. The incorporation of an ancillary
system considerably reduces the complexity of the preparation stage
(as opposed to exponentially many input states in SQPT, only one state
has to be prepared and only one output state has to be characterized)
but comes at the cost of needing access to an independent ancilla
system of the same size as the system that is to be analyzed. Still,
for a complete characterization of the output state
$\hat{\varrho}_{\mathcal{E}}$, a number of measurements exponentially
large in $n$ is required \cite{Mohseni2008}.

In this work, we combine recent results from efficient and scalable
state tomography \cite{Cramer2010,Baumgratz2013a} with the idea of
AAPT and discuss how to avoid the need for an ancilla system (see,
e.g., \cite{Flammia2012,Bendersky2008}) as summarized in
Fig.~\ref{fig:measurement-scheme}.  This allows us to formulate a
scalable process tomography scheme where the number of input and
output states, together with the experimental measurement settings to
characterize the latter, grows only linearly with the number of
qubits.

We restrict our attention to unitary processes and demonstrate our
scalable process tomography technique with numerical simulations of
quantum circuits and of time evolution under local one-dimensional
Hamiltonians. In Sec.~\ref{sec:numerics-circuits}, we consider a
circuit which prepares a Greenberg-Horne-Zeilinger (GHZ) state and the
quantum Fourier transform. Note that the GHZ circuit is a
limited-depth circuit (as defined below, cf.~\cite{Jozsa2006}), which
implies that it has an efficient matrix product operator (MPO)
representation in the sense that the bond dimension scales at most
polynomially with the number of qubits. This is exactly the class of
processes for which one would expect methods based on
Refs.~\cite{Cramer2010,Baumgratz2013a} to work. In turn, the quantum
Fourier transform is a circuit of polynomial depth, but one that
admits an approximate efficient simulation with a classical
computer~\cite{Yoran2007}. For the $n=32$ qubits considered here, an
MPO representation that approximates the circuit exists and our
reconstruction scheme works well.

In Sec.~\ref{sec:numerics-hamiltonians} we simulate process tomography
for time evolutions of one-dimensional local Hamiltonians $\hat H$ and
show how to extract the Hamiltonians from the reconstructed
processes. The reconstruction of a Hamiltonian on $n$ qubits will
require one process reconstruction at time
$\sim 1/\|\hat H\| \sim 1/n$
(Sec.~\ref{sec:numerics-hamiltonians-essentials}) or two process
reconstructions at times $t_{1}$, $t_{2}$ separated by no more than
$\sim 1/\|\hat H\|$
(Sec.~\ref{sec:numerics-hamiltonians-twotimes}). In the latter case,
the only restriction on $t_{1}$ and $t_{2}$ is that the process admits
reconstruction at those times. In both cases, the reconstruction of a
Hamiltonian will require roughly $\sim\!n^{2}$ measurements of each of
the $\sim\!n$ observables.

The quality of the reconstructed processes may be quantified using the
same experimental resources that are also required to obtain the
reconstruction: For unitary processes, $\hat{\varrho}_{\mathcal{E}}$
is guaranteed to be pure such that certifiability of the
reconstruction of $\mathcal{E}$ is inherited by the certifiablity of
$\hat{\varrho}_{\mathcal{E}}$ \cite{Cramer2010}; see
Ref.~\cite{Kim2014} for an analogous mixed-state certificate.

\section{Scalable Process Tomography}

Let us first consider the AAPT scheme and restrict, without loss of
generality, to qubits. Furthermore, we arrange the ancilla + system as
depicted in Fig.~\ref{fig:measurement-scheme} with odd sites
representing the ancilla and even sites denoting the system itself. In
this enumeration, the maximally entangled input state $|\Phi\rangle$
takes the form of a product of Bell states
$|\phi^{+}\rangle_{1,2}\otimes\cdots |\phi^{+}\rangle_{2n-1,2n}$,
$|\phi^{+}\rangle \propto |00\rangle + |11\rangle$, and hence only
requires local two-qubit manipulations for its experimental
generation.  The system (i.e., the even sites) is sent through the
channel and state tomography has to be performed on the resulting
state $\hat{\varrho}_{\mathcal E}$. Without any prior knowledge about
the underlying quantum channel, full state tomography is inevitable
and one seemingly faces the notorious curse of dimensionality. A large
class of quantum states, however, may be reconstructed from a number
of measurements and with post-processing resources that both scale
only polynomially in $n$
\cite{Cramer2010,Baumgratz2013,Baumgratz2013a}.

Post-processing with polynomial resources requires an efficient
representation of the state, which, in one dimension, is provided by
matrix product state (MPS) or operator (MPO) representations
\cite{Fannes1992,Perez2006,Schollwoeck2011}, compare also tree tensor
networks \cite{Shi2006} and the multiscale entanglement
renormalization ansatz \cite{Vidal2007}. There are many examples of
physical states that admit an efficient matrix product representation,
e.g. ground states of gapped local Hamiltonians
\cite{Hastings2007,Plenio2005,Eisert2010}, thermal states of local
Hamiltonians \cite{Eisert2010,Hastings2006} and entangled states like
the GHZ state, the W state and cluster states.

If the output state $\hat{\varrho}_{\mathcal E}$ happens to be close
to this class of states with an efficient matrix product
representation, scalable reconstruction of the channel may be
achieved.  The input to the scalable state tomography schemes
\cite{Cramer2010,Baumgratz2013,Baumgratz2013a} are given by the
measurement data of the following observables (or, alternatively, any
other local operator basis):
\begin{equation}
  \label{Eq:observables}
  \hat{P}_{k;\alpha_1,\dots,\alpha_r}=
  \id_{1,\dots,k}\otimes \hat{\sigma}^{\alpha_1}_{k+1}\otimes\cdots\otimes \hat{\sigma}^{\alpha_r}_{k+r}\otimes\id_{k+r+1,\dots,2n},
\end{equation}
for all $\alpha_i\in\{x,y,z\}$, $k=0,\dots,2n-r$, and a fixed $r$
independent of the size of the system. Here, $\hat{\sigma}_i^x$,
$\hat{\sigma}_i^y$, $\hat{\sigma}_i^z$ are the Pauli matrices for
qubit $i$. There are $(2n-r+1)\times 3^r$ such operators, i.e., the
number of observables that are required for these reconstruction
schemes scales linearly in $n$. Note that the restriction to local
information is not mandatory, any output state that is uniquely
characterized by a number of measurements that scales moderately with
$n$ can be reconstructed by these techniques \cite{Baumgratz2013a}
and, hence, belongs to the class of states for which our process
tomography procedure is applicable.

Interestingly, the necessary information may also be obtained without
the need of an ancilla (see, e.g., \cite{Flammia2012,Bendersky2008}),
yet then increasing the demand at the preparation stage: By virtue of
the identity
\begin{equation}	
  \text{tr}[(\hat{A}\otimes \hat{S})\hat{\varrho}_{\mathcal E}]=\frac{\text{tr}[\mathcal E(\hat{A}^t)\hat{S}]}{2^{n}} ,\;\;\;\hat{A}^t=\sum_{i,j}\langle j|\hat{A}|i\rangle |i\rangle\langle j|,
\end{equation}
which holds for any operator $\hat{S}$ ($\hat{A}$) acting on the
system (ancilla), and the fact that each
$\hat{P}_{k;\alpha_1,\dots,\alpha_r}$ is of the form
$\hat{P}_A\otimes \hat{P}_S$ ($A$: ancilla, $S$: system), one may
obtain the necessary expectation values by preparing the eigenstates
of the Pauli matrices in $\hat{P}_A$ on the system, sending them
through the channel and measuring $\hat{P}_S$ on the resulting state
(see Fig.~\ref{fig:measurement-scheme} and Appendix\
\ref{appendix:equivalence_details} for details) -- a scheme that
requires no ancilla, the preparation of linearly many states and the 
measurement of linearly many observables.

While the preparation and measurement strategy we just outlined may be
favourable from an experimental perspective, we will present our
scheme in the framework of AAPT as certain intuitions are particularly
transparent in this setting. In the remainder of this section, we
discuss how unitary operators and the Hamiltonians governing time
evolution can be constructed from $\hat\varrho_{\mathcal{E}}$. In
principle, the scheme is applicable to non-unitary channels as well,
as long as the corresponding state $\hat \varrho_{\mathcal E}$ permits
reconstruction.

\subsection{Reconstruction of Unitary Channels\label{sec:reconstruct-unitaries}}

We aim at reconstructing unitary channels, so channels of the form
$\mathcal{E}(\hat{\varrho})=\hat{U}\hat{\varrho }\hat{U}^\dagger$. For
those, it is guaranteed that the resulting state is pure, i.e.,
$\hat{\varrho}_{\mathcal{E}}=|\psi_{\mathcal{E}}\rangle\langle\psi_{\mathcal{E}}|$.
The unitary may then be obtained from the identity
$\langle j|\hat{U}|i\rangle=2^{n/2}\langle i|\langle
j|\psi_{\mathcal{E}}\rangle$.
The output of the state reconstruction algorithms of
Refs.~\cite{Cramer2010, Baumgratz2013a} provide us with a pure
estimate $|\psi^\text{rec}_{\mathcal{E}}\rangle$ of
$\hat{\varrho}_{\mathcal{E}}$ given in the form of a matrix product
state (MPS) \cite{Fannes1992,Perez2006},
\begin{equation}
  \ket{\psi^{\text{rec}}_{\mathcal{E}}} =
  \sum_{i_{1},\ldots,i_{2n}}
  A^{(1)}_{i_{1}} \cdots A^{(2n)}_{i_{2n}}
  \ket{i_{1}, \ldots ,i_{2n}}, 
\end{equation}
where $A^{(k)}_{i_{k}} \in \mathbb C^{D_{k-1} \times D_{k}}$ with
$D_{0} = 1 = D_{2n}$ and summation is over all $i_k=1,2$;
$k=1,\ldots,2n$.  The $D_{k}$ are called bond dimensions. Given this
form, the matrix product operator (MPO) representation of the estimate
$\hat{U}_\text{rec}$ to $\hat{U}$,
$\langle j|\hat{U}_\text{rec}|i\rangle=2^{n/2}\langle i|\langle
j|\psi^\text{rec}_{\mathcal{E}}\rangle$,
may straightforwardly be obtained by grouping the $n$ pairs of ancilla
and system sites:
\begin{widetext}
  \begin{equation}
    \label{eqn:MPO}
    \hat{U}_{\text{rec}} =
    2^{\frac n2}
    \sum_{i_{1},\ldots,i_{2n}} 
    \Bigl[A_{i_{1}}^{(1)} A_{i_{2}}^{(2)}\Bigr] \cdots 
    \Bigl[A_{i_{2n-1}}^{(2n-1)} A_{i_{2n}}^{(2n)}\Bigr]\,\,
    \ket{i_{2}}\!\bra{i_{1}}
    \otimes \cdots \otimes
    \ket{i_{2n}}\!\bra{i_{2n-1}}.
  \end{equation}
\end{widetext}
As the input state $|\Phi\rangle$ can efficiently be represented as an
MPS (given the sites are labeled as depicted in
Fig.~\ref{fig:measurement-scheme}), the process $\hat U$ having an
efficient MPO representation will imply that the exact state
$\ket{\psi_{\mathcal E}}$ has an efficient MPS representation, and
successful reconstruction may be possible.
We now show that circuits with an at most logarithmic depth have an
efficient MPO representation. Our condition is equivalent to the
condition given in Ref.~\cite{Jozsa2006}, but we give a tighter bound
on the bond dimension.  Let a circuit $\hat U$ be composed of $N$
two-qubit gates,
\begin{align*}
  \hat U = \prod_{j=1}^{N} \hat U_{j}, \quad
  \hat U_{j} \text{ acts on } l_{j} \text{ and } r_{j} > l_{j}.
\end{align*}
We define the depth of the circuit at the bipartition $i|i+1$ by
\begin{align*}
  d_{i} = |\{ j \colon l_{j} \le i \text{ and } r_{j} \ge i+1 \}|
\end{align*}
and denote the maximal depth by $d_{\text{max}} = \max_{i} d_{i}$.
This definition is motivated by the fact that an MPO representation of
$\hat U$ with bond dimension at most $4^{d_{\text{max}}}$ is easily
obtained from the MPO representations of the $\hat U_{j}$: The
$\hat U_{j}$ have bond dimension $D_{i} \le 4$ for
$l_{j} \le i \le r_{j}-1$ and $D_{i} = 1$ everywhere else, and the
bond dimension of the product $\hat U$ is given, in the worst case, by
$D_{i}(\hat U) = \prod_{j=1}^{N} D_{i}(\hat U_{j})$
\cite{Schollwoeck2011}.
If the depth $d_{\text{max}}$ grows at most logarithmically with~$n$,
we obtain an MPO representation of the circuit that is efficient in
the sense that it has a number of parameters at most polynomial in
$n$.

\subsection{Hamiltonian Reconstruction\label{sec:reconstruct-hamiltonians}}

Assume that the quantum process is in fact the time evolution under a
local one-dimensional Hamiltonian, i.e.,
$\hat{U}=\e^{-\ii \hat{H} t}$, with $\hat{H} = \sum_{i} \hat{h}_{i}$,
where $\hat{h}_{i}$ only acts on a fixed number of neighbouring sites
(we will consider nearest-neighbour Hamiltonians throughout) and where
$\|\hat h_{i}\| \le J$ for a constant $J$. With the tools to
reconstruct unitary processes at hand, it remains to address the
question of how to find a valid estimate of the Hamiltonian $\hat{H}$
governing the time evolution. To obtain $\hat{H}$ from this unitary,
we will use the identity
\begin{align}
  x &= \sin(x) \frac{\arccos(\cos(x))}{\sqrt{1-(\cos(x))^{2}}}, & 
  x &\in (-\pi, \pi),
  \label{eq:trigon-sin-cos-arccos}
\end{align}
together with the power series 
\begin{align}
  \frac{\arccos(z)}{\sqrt{1-z^{2}}} &= \sum_{k=0}^{\infty} c_{k}
  (z-1)^{k}, & c_{k} = \frac{(-1)^{k}}{2^{k}} \prod_{j=1}^{k}
  \frac{j}{j+\frac12} 
  \label{eq:arccos-series}
\end{align}
which converges for $|z-1| < 2$ %
\cite{[][{{. Print compantion to: NIST Digital Library of Mathematical
      Functions. \unexpanded{\url{http://dlmf.nist.gov/}}, Release
      1.0.9 of 2014-08-29. Derived from Eq.~4.24.2, available at
      \unexpanded{\url{http://dlmf.nist.gov/4.24\#E2}}.}}]Olver2010nocompanion}.  %
The %
basic idea is that from $\hat{U} = \e^{-\ii \hat{H} t}$, we know that
$2\ii\sin(\hat{H} t) = \hat{U}^{\dagger} - \hat{U}$,
$2\cos(\hat{H} t) = \hat{U}^{\dagger} + \hat{U}$, and Eq.\
(\ref{eq:trigon-sin-cos-arccos}) holds up to times limited by
$\| \hat{H} t \| < \pi$.  While this appears to limit the accessible
time interval for Hamiltonian reconstruction,
Sec.~\ref{sec:numerics-hamiltonians-twotimes} will explain how to
extend this result to longer times.

For practical purposes, we only want to evaluate a finite number of
terms of the series in Eq.~\eqref{eq:arccos-series} and enforce
Hermiticity. To this end, we approximate
\begin{equation}
  \label{eqn:RecHamiltonian}
  \hat{H}_\text{rec}t =\frac{1}{2}
  \sin(\hat{H}_{\text{rec}}t) \sum_{k=0}^{N-1} c_{k} (\cos(\hat{H}_{\text{rec}}t)-\id)^{k}+\text{h.c.}
\end{equation}
for a given $N$ and $\sin(\hat{H}_{\text{rec}} t)$ and
$\cos(\hat{H}_{\text{rec}}t)$ as above. This approach has two
advantages over using a power series expansion of the logarithm: It is
valid for larger values of $\|Ht\|$ and the series converges much
faster. Note that even if the reconstruction of
$|\psi_{\mathcal{E}}\rangle$ is perfect, $\hat{U}_\text{rec}$ may
differ from $\hat{U}$ by a global phase $\phi$, such that
$\hat{U}_\text{rec}^{\dagger} \e^{\ii\phi} \hat{U}=\id$ and Eq.\
\eqref{eq:trigon-sin-cos-arccos} imposes $\| Ht - \phi \id \| < \pi$.
To remedy this problem, we use
$\hat{U}_\text{rec}
\tr(\hat{U}_\text{rec})^{*}/|\tr(\hat{U}_\text{rec})|$
as our actual estimate.

The initial state $|\Phi\rangle$ is an MPS of low bond dimension and,
as under local Hamiltonians quantum correlations build up in a
light-cone-like picture, $\ket{\psi_{\mathcal{E}}}$ will still have a
small bond dimension for small times \cite{Eisert2006,
  Bravyi2006}. More precisely, at a fixed time, there is an
approximate MPO representation of $U$ with a bond dimension that grows
at most polynomially with the number of qubits \cite{Osborne2006}.  In
our numerical simulations on $n \le 32$ qubits, we observe that a MPS
representation is feasible at times on the order of $1/J$.

\section{Numerical Simulations}

We carry out numerical simulations as follows: For several exemplary
channels $\mathcal{E}$, we numerically obtain
$\ket{\psi_{\mathcal E}}$ as detailed below and simulate measurements
of the local observables $\hat{P}_{k;\alpha_1,\dots,\alpha_r}$ by
drawing $M$ times per observable according to
$\ket{\psi_{\mathcal E}}$. In this way, we take statistical errors
into account, with statistical errors for measurements without ancilla
being very similar \cite{ft:noancilla-statistical-errors}.
The resulting empirical mean values (i.e., the simulated estimates of
the weights
$\langle \hat{\Pi}_{k;\alpha_1,\ldots,\alpha_{r}}^{s_{1},\ldots,s_{r}}
\rangle_{\hat{\varrho}_{\mathcal{E}}}$
of all eigenprojectors
$\hat{\Pi}_{k;\alpha_1,\ldots,\alpha_{r}}^{s_{1},\ldots,s_{r}}$,
$s_i=\pm 1$, of $\hat{P}_{k;\alpha_{1},\ldots,\alpha_{r}}$) are then
used to obtain a reconstruction
$|\psi_{\mathcal{E}}^\text{rec}\rangle$ of $\ket{\psi_{\mathcal E}}$.
This is done by exploiting the singular-value-thresholding-like
algorithm described in \cite{Cramer2010} to obtain an initial state
for the scalable maximum-likelihood algorithm of Ref.\
\cite{Baumgratz2013a}. The result is an MPS representation of
$|\psi_{\mathcal{E}}^\text{rec}\rangle$ which can be converted into an
MPO representation of $\hat{U}_\text{rec}$ as described above in Eq.\
(\ref{eqn:MPO}).

With the estimate $|\psi_{\mathcal{E}}^\text{rec}\rangle$ and thus the
corresponding operator $\hat{U}_\text{rec}$ at hand, we then quantify
the quality of the reconstruction scheme by
\begin{equation}
  \label{eq:u-fidelity}
  F=F(\hat{U},\hat{U}_\text{rec})=|\langle \psi_{\mathcal{E}}|\psi^\text{rec}_{\mathcal{E}}\rangle|^2,
\end{equation}
and note that this is in one-to-one correspondence to other distance
measures for unitary channels used in the literature
\cite{Raginsky2001, Gilchrist2005, Pedersen2007,
  ft:process-fidelity-relations}.

In the case of Hamiltonian reconstruction, we assess our reconstructed
estimates as follows: First, note that two Hamiltonians $\hat{H}$ and
$\hat{H} + \lambda \id$, $\lambda \in \mathbb R$, are physically
indistinguishable. Therefore, we measure relative distances between
Hamiltonians according to \cite{ft:dmrg-ground-state-search}
\begin{align}
  D(\hat{H}, \hat{H}^{\prime}) &= \frac{\min_{\lambda \in \mathbb R} \| \hat{H} - \hat{H}^{\prime} - \lambda
    \id \|} {\min_{\lambda \in \mathbb R} \| \hat{H} - \lambda \id
    \|},
  \label{eq:hamiltonian-distance}
\end{align}
which is independent of energy offsets in both $\hat{H}^{\prime}$ and
$\hat{H}$.  We choose the operator norm $\|\cdot\|$ motivated by its
property
\begin{align*}
  | \langle \hat{A}(t) \rangle_{\hat{\varrho}} - \langle \hat{A}^{\prime}(t) \rangle_{\hat{\varrho}} | \le 2 |t| \| \hat{H} - \hat{H}^{\prime} \| \| \hat{A} \|,
\end{align*}
where $\hat{A}(t)$ and $\hat{A}^{\prime}(t)$ are the Heisenberg
picture time evolutions of $\hat{A}$ according to $\hat{H}$ and
$\hat{H}^{\prime}$, respectively. In other words, the operator norm
distance defines a timescale on which two Hamiltonians may be
considered equivalent.

For all results below, we repeat the whole procedure of sampling from
the simulated state, reconstructing it, and assessing the quality of
the reconstruction several times. All results shown are mean values of
$F(\hat{U},\hat{U}_\text{rec})$ and $D(\hat{H},\hat{H}_\text{rec})$
over a small number of runs, with deviations that are, for the number
of measurements per observable considered, smaller than the size of
the markers. Next, we present numerical results for the reconstruction
of quantum circuits and Hamiltonians and study the performance as a
function of the number of qubits $n$, the number $M$ of measurements
per observable, and the block size $r$ of the subsystems on which
measurements are performed. We simulate circuits and Hamiltonians on
up to $32$ qubits. Hence, reconstructing the unitary uses pure state
reconstruction on up to $64$ qubits.

\subsection{Quantum Circuits \label{sec:numerics-circuits}}

We demonstrate the feasibility of our scalable tomography scheme by
considering the $\widehat{\text{GHZ}}$ circuit, which prepares an
$n$-qubit GHZ state from $\ket{0 \ldots 0}$ and the quantum Fourier
transform \cite{Nielsen2007},
\begin{equation}
  \label{eq:circuits}
  \begin{split}
    \widehat{\text{GHZ}} &= \widehat{\text{CN}}_{n-1,n}
    \widehat{\text{CN}}_{n-2,n-1} \cdots
    \widehat{\text{CN}}_{1,2} \hat{H}_{1},\\
    \widehat{\text{QFT}} &= \prod_{k=1}^{n} \Biggl[ \Biggl(
    \prod_{j=1}^{n-k} \widehat{\text{CR}}_{k,k+j}(\pi/2^{j}) \Biggr)
    \hat{H}_{k} \Biggr],
  \end{split}
\end{equation}
where we use the convention
$\prod_{j=1}^{k}\hat{U}_i = \hat{U}_{k}\cdots \hat{U}_{1}$ for
products of non-commuting operators. Here, $\hat{H}_k$ denotes the
Hadamard gate acting on qubit $k$, and $\widehat{\text{CN}}_{i,i+1}$
($\widehat{\text{CR}}_{i,j}(\phi)$) denotes the two-qubit
conditional-NOT (conditional rotation) gate \cite{Nielsen2007}.
The $\widehat{\operatorname{GHZ}}$ circuit has depth
$d_{\text{max}} = 1$ and thus admits an exact efficient MPO
representation. The depth of the exact quantum Fourier transform
circuit is $d_{\text{max}} = \lfloor \frac{n^{2}}{4} \rfloor$ and
grows quadratically with the number of qubits (see
Appendix~\ref{appendix:qft-properties}).  However, one can obtain an
approximation $\widehat{\operatorname{QFT}}_{c}$ of the quantum
Fourier transform by dropping all conditional rotations with $j > c$
from the definition in Eq.~\eqref{eq:circuits}
\cite{Coppersmith1994}. This approximation can be simulated
classically with polynomial resources \cite{Yoran2007}. Using
numerical MPO compression techniques~\cite{Schollwoeck2011}, we obtain
an approximate MPO representation with bond dimension 16 and error
bounded by $[2(1-\sqrt F)]^{1/2} < 2 \times 10^{-5}$ for the
$n \le 32$ qubits we consider.

\begin{figure}[t]
  \begin{center}
    \includegraphics[width=0.99\columnwidth]{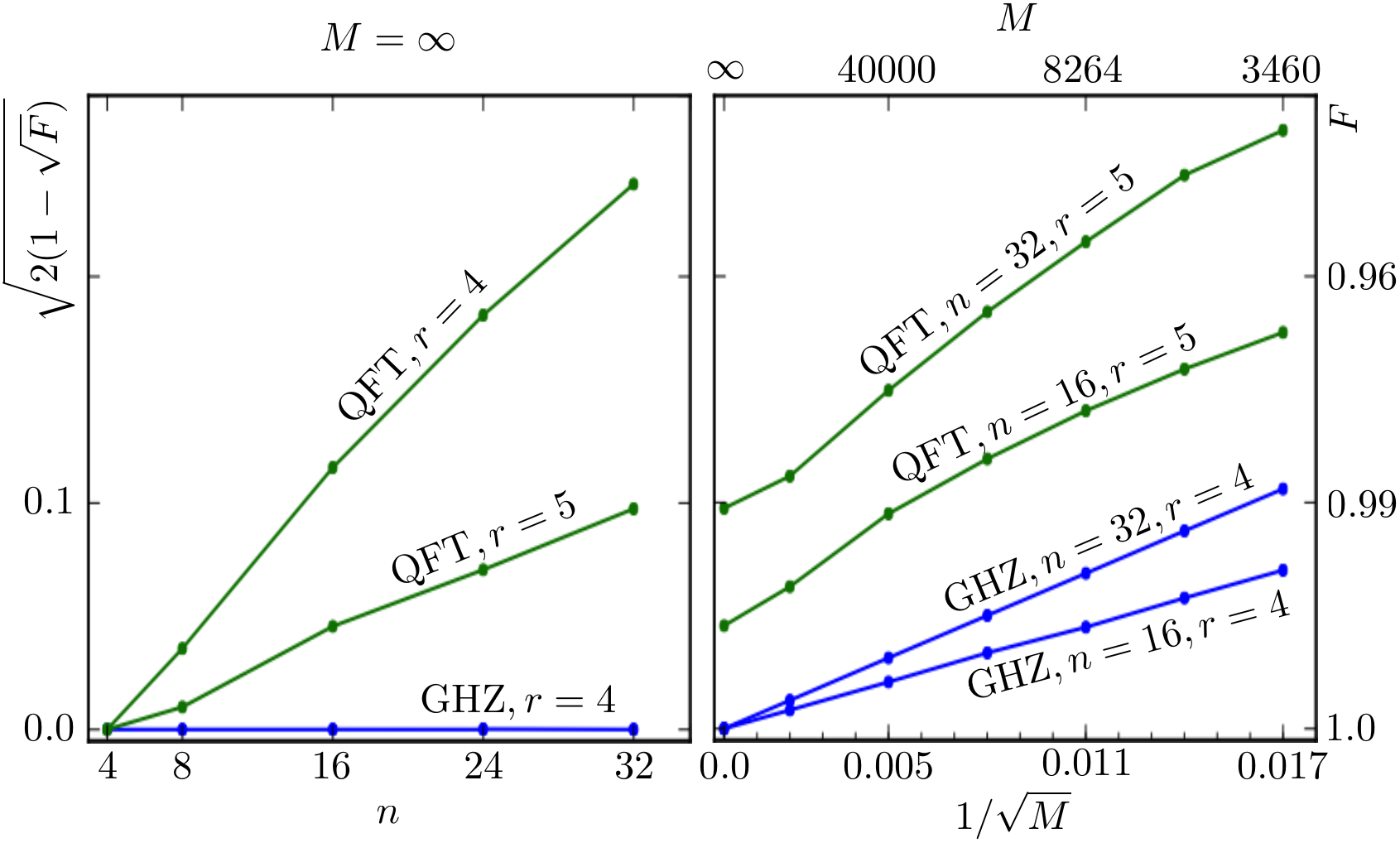}\vspace{-0.3cm}
  \end{center}
  \caption{(Color online) Reconstruction of the quantum circuits in
    Eq.\ \eqref{eq:circuits}. Figures show the fidelity in
    Eq.~\eqref{eq:u-fidelity} of the exact unitary and its tomographic
    reconstruction as a function of the system size $n$ for $M=\infty$
    (left) and as a function of the number $M$ of measurements per
    local observable for different $n$ (right).  Reconstruction uses
    measurements on all blocks of $r$ consecutive qubits only.  Lines
    are guides to the eye. The numerical data suggest a linear scaling
    of $(1-\sqrt{F})^{1/2}$ with system size $n$ and $1/\sqrt{M}$.  }
     \label{fig:gate-reconstruction}
\end{figure}

The reconstruction results are summarized in
Fig.~\ref{fig:gate-reconstruction}.  The reconstruction of the
$\widehat{\operatorname{GHZ}}$ performs very well. To discuss the
performance of the quantum Fourier transform reconstruction, we note
that the distance $\epsilon_{c}$ between the exact quantum Fourier
transform and its approximation $\widehat{\operatorname{QFT}}_{c}$ is
upper bounded by $\epsilon_{c} \le n \pi / 2^{c}$ (see
\cite{Yoran2007} and Appendix~\ref{appendix:qft-properties}).
We reconstruct $\widehat{\operatorname{QFT}}$ with high fidelity
$F \approx 0.99$ from measurements on $r=5$ consecutive qubits on the
combined system + ancilla (Fig.~\ref{fig:measurement-scheme}). Naively,
one would expect to be able to reconstruct
$\widehat{\operatorname{QFT}}_{c}$ only for $c \le 2$, because $r=5$
corresponds to information about three neighbouring system qubits
only. However, the upper bound on the approximation error is trivial
for $c = 2$ and $8 \le n \le 32$, and numerical tests show that the
actual approximation error $\epsilon_{c}$ is indeed several times
larger than the reconstruction error $[2(1-\sqrt F)]^{1/2}$ we
achieve. This shows that there are non-local gates which can be
reconstructed without using the corresponding non-local information.

\subsection{Hamiltonian reconstruction\label{sec:numerics-hamiltonians}}

\subsubsection{Short times\label{sec:numerics-hamiltonians-essentials}}

\begin{figure*}[t]
  \centering
  \includegraphics[width=\linewidth]{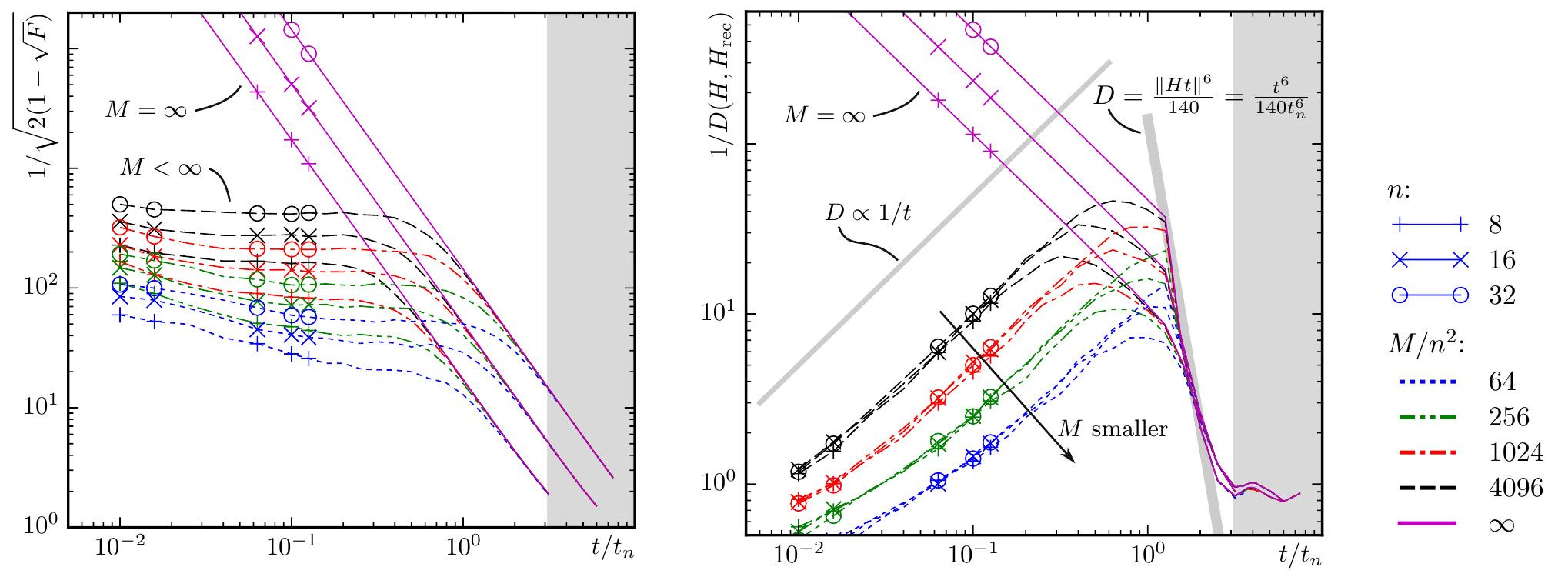}
  \caption{(Color online) \textit{Left:} Fidelity between exact state
    $\ket{\psi_{\mathcal E}}$ and tomographic estimate
    $\ket{\psi_{\mathcal E}^\text{rec}}$ of a unitary time evolution
    $\hat{U} = \e^{-\ii \hat{H} t}$, with $\hat{H}$ the isotropic
    Heisenberg Hamiltonian on $n$ qubits. The tomographic estimate is
    based on complete measurements on blocks of $r = 3$ consecutive
    qubits with $M$ measurements per observable.  \textit{Right:}
    Inverse relative distance~$D$ between $\hat{H}$ and
    $\hat{H}_\text{rec}$ reconstructed from
    $\ket{\psi_{\mathcal E}^\text{rec}}$.  The unit of time is
    $t_{n} = 1/\|\hat{H}\| \sim 1/n$, i.e., inversely proportional to
    the number of qubits, and the grey area indicates where
    reconstruction is expected to fail due to
    Eq.~\eqref{eq:trigon-sin-cos-arccos}. The grey lines without any
    markers show the expected behaviour for small times and times near
    $\pi$ (see Eq.~\eqref{eq:hrec-series-error-estimate} and the main
    text).  The data indicates that for $M < \infty$, $M/n^{2}$ fixed
    and $t/t_{n}$ fixed and sufficiently small,
    $D(\hat{H},\hat{H}_{\text{rec}})$ is largely independent of $n$.
    }
  \label{fig:hamiltonian-reconstruction}
\end{figure*}

We simulate the time evolution $\hat U$ of time-independent local
one-dimensional Hamiltonians $\hat H$ with well-established numerical
DMRG/MPO algorithms~\cite{ft:dmrg-time-evolution}. After obtaining the
estimate $\hat U_\text{rec}$ of the time evolution, we determine an
estimate $\hat H_\text{rec}$ of the Hamiltonian that governs the time
evolution by the series given in Eq.~\eqref{eqn:RecHamiltonian} with
$N = 3$. With this, and the assumption that
$\min_{\lambda \in \mathbb R} \| \hat{H} - \lambda \id \| = \| \hat{H}
\|$ \cite{ft:h-rec-scaling-prefactor}, one has
\begin{align}
  D(\hat{H}, \hat{H}_\text{rec}) & \le
  \tfrac{\| \hat{H} - \hat{H}_{\text{rec}} \|}{\| \hat{H} \|}
  \le
  \tfrac{1}{140}
  \| \hat{H}t \|^{6} + O(\| \hat{H}t \|^{8}).
  \label{eq:hrec-series-error-estimate}
\end{align}
Fig.~\ref{fig:hamiltonian-reconstruction} shows results for an
isotropic Heisenberg Hamiltonian on $n \le 32$ qubits. We have also
studied the critical Ising model and a Hamiltonian with random
nearest-neighbour interaction, which show very similar behaviour and
the corresponding results may be found in Appendix~\ref{appendix:other
  hamiltonians}. We use $t_{n} = 1/\| \hat H\|$ as a time unit and
recall that the local terms of the Hamiltonian are bounded by a
constant, $\|\hat h_{i}\| \le J$. This gives us $t_{n} \ge 1/n J$ and
$t_{n} \sim 1/n$ if we additionally assume that the local terms all
have similar norm.

The distance $D(\hat{H}, \hat{H}_\text{rec})$ between the
reconstructed and the exact Hamiltonian shown in
Fig.~\ref{fig:hamiltonian-reconstruction} displays the following
features: First, the reconstruction is expected to fail for
$\| \hat{H} t \| \ge \pi$ (see Eq.~\eqref{eq:trigon-sin-cos-arccos})
and, indeed, $D(\hat{H}, \hat{H}_\text{rec})$ is large in this area
(indicated by the grey background).  Secondly,
Eq.~\eqref{eq:hrec-series-error-estimate} suggests that close to
$\| \hat{H} t \| = \pi$, $D(\hat{H}, \hat{H}_\text{rec})$ should scale
as $ \| \hat{H}t \|^{6}/140=t^6/(140t_n^6)$ (thick grey line in
Fig.~\ref{fig:hamiltonian-reconstruction}).  Thirdly, we observe that
for infinitely many measurements per observable, $M = \infty$, and
fixed $t/t_n=\|\hat{H}\|t\sim nt$, the distance
$D(\hat{H}, \hat{H}_\text{rec})$ decreases with system size, a
behaviour inherited from the quality of the reconstruction
$|\psi^\text{rec}_{\mathcal E}\rangle$ (see left of
Fig.~\ref{fig:hamiltonian-reconstruction}): The fidelity
$F(\hat{U},\hat{U}_\text{rec})$ is limited by the amount of block
entanglement in $|\psi_{\mathcal{E}}\rangle$. At a fixed time $t$, an
area law \cite{Eisert2010} holds for this entanglement such that it is
bounded even for arbitrarily large systems \cite{Eisert2006,
  Bravyi2006}. For sufficiently large systems, we hence expect
$F(\hat{U},\hat{U}_\text{rec})$ at fixed $t$ to be independent of the
system size $n$. If we keep $t/t_n \sim t \cdot n$ fixed, we therefore
expect $F$ to increase with $n$. Finally, let us discuss the
dependence of the distance between the exact and the reconstructed
Hamiltonian in Fig.~\ref{fig:hamiltonian-reconstruction} on the number
$M$ of measurements per observable. First of all, with a finite number
of measurements no reconstruction will be possible at small times,
because the signal of the Hamiltonian in
$\hat{U} \approx \id - \ii \hat{H} t$ will be smaller than the
noise. Furthermore, the data suggests that, for times before
$t/t_{n} \approx \pi$,
\begin{align}
D(\hat{H}, \hat{H}_\text{rec})
  \propto &
  \frac{1}{t/t_{n}} \frac{n}{\sqrt M}.
\label{eq:number-of-measurements-scaling}
\end{align}
This is the behaviour one would expect if one assumes that the
relative error $D(\hat{H}, \hat{H}_\text{rec})$ is proportional to the
ratio $R/S$ of a noise amplitude $R$ and the strength of the signal
$S = \| \hat{H} t \|=t/t_n$, in which $R = n / \sqrt M$ is motivated
by the fact that we have measured $\propto n$ observables, each of
which has been estimated to within a standard deviation given, for
sufficiently large $M$, by $1/\sqrt M$.

The explanation of the scaling properties together with the fact that
all properties described are the same in the other Hamiltonians
investigated (see Appendix~\ref{appendix:other hamiltonians}) suggest
that that the scaling laws apply for many local Hamiltonians on a
linear chain, for a large range of system sizes and any sufficiently
large (as indicated by the examples) number of measurements.

To summarize, measuring at larger times gives a larger signal and a
smaller error, but we are limited by the condition $ t/t_{n} < \pi$
imposed by Eq.~\eqref{eq:trigon-sin-cos-arccos}.  Solving
Eq.~\eqref{eq:number-of-measurements-scaling} for the number of
measurements per observable, we obtain
$M \propto n^{2} (t/t_{n})^{2} / D^{2}$: A constant relative error $D$
at a fixed $t/t_{n} < \pi$ requires $M \propto n^{2}$ measurements per
observable, resulting in a total number of measurements proportional
to $n^{3}$ .

\begin{figure*}[t]
  \centering
  \includegraphics{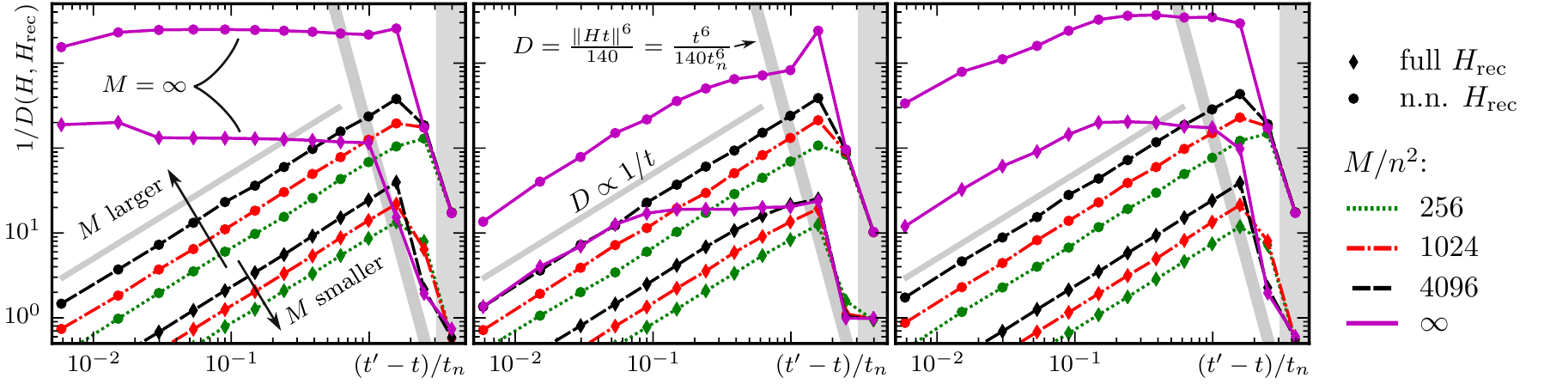}
  \caption{(Color online) Quality of the reconstruction
    $\hat{H}_\text{rec}$ for $\hat{H}$ the critical Ising (left),
    isotropic Heisenberg (middle) and a random nearest-neighbour
    Hamiltonian (right).  $\hat{H}_\text{rec}$ was obtained from
    $U(t'-t)$ with $t/t_{n} = 3.51$ fixed and
    $U(t'-t) = U^{\dagger} U'$ was obtained from tomographic estimates
    of $U = \e^{-\ii \hat H t}$ and $U' = \e^{-\ii \hat H t'}$.  The
    inverse relative error $1/D(\hat{H},\hat{H}_{\text{rec}})$ is
    shown for $H_\text{rec}$ (diamonds) and for the projection of
    $H_\text{rec}$ onto a nearest-neighbour Hamiltonian (circles).
    $\hat H$ acts on $n = 32$ qubits and tomographic estimates are
    based on complete measurements on blocks of $r = 5$ consecutive
    qubits with $M$ measurements per observable. The unit of time is
    $t_{n} = 1/\|\hat{H}\| \sim 1/n$.  Most properties are similar to
    Fig.~\ref{fig:hamiltonian-reconstruction}~right where
    $\hat H_\text{rec}$ was reconstructed from an estimate of $U(t)$,
    with the time difference $t'-t$ taking the role of the time
    $t$. Lines are guides to the eye.  }
  \label{fig:hamiltonian-reconstruction-twotimes}
\end{figure*}

\subsubsection{{Long times}\label{sec:numerics-hamiltonians-twotimes}}

In its present formulation, the reconstruction scheme is limited to
$t/t_n<\pi$, a restriction that may be overcome by measuring at two
different times $t$, $t^\prime$: The times up to which the fidelity
$F(\hat{U},\hat{U}_\text{rec})$ is sufficiently high is only limited
by $r$ -- increasing $r$ will increase the time up to which full
information about $\hat{U}$ may be obtained by measuring on $r$
consecutive qubits. In fact, as can be seen on the left of
Fig.~\ref{fig:hamiltonian-reconstruction}, for $n = 32$ and the
relatively small $r=3$, the fidelity $F(\hat{U},\hat{U}_\text{rec})$
is still quite high at $t/t_{n} = \pi$ while the reconstruction of
$\hat{H}$ fails for these times.  Measuring at $t$, $t^\prime$ and
obtaining $\hat{U}=\e^{-\ii \hat{H}t}$,
$\hat{U}^\prime=\e^{-\ii \hat{H}t^\prime}$ by reconstruction, we are
only limited by $|t^\prime-t|<\pi t_n$ when reconstructing $\hat{H}$
from $\hat{U}^\dagger \hat{U}^\prime=\e^{\ii \hat{H}(t-t^\prime)}$.
Fig.~\ref{fig:hamiltonian-reconstruction-twotimes} shows results of
this reconstruction scheme with $t/t_{n} = 3.51$ and $t' > t$.

Reconstructing the Hamiltonian from
$\hat U(t'-t) = \hat U^{\dagger}\hat U'$, the time difference $t'-t$
clearly assumes the role of the time $t$ when reconstructing the
Hamiltonian from $\hat U$ at time $t$ alone. Therefore, all scaling
properties carry over as long as $\hat U_\text{rec}$ and
$\hat U'_\text{rec}$ can be obtained with sufficiently high
fidelity. We simulated measurements on blocks of $r = 5$ consecutive
sites to satisfy this requirement.

For exact measurements, $M = \infty$, the relative error
$D(\hat H, \hat H_\text{rec})$ does not approach zero for
$t'-t \to 0$.  The reason is that the error in
$\hat U_\text{rec}(t'-t)$ remains non-zero as $t'-t\to 0$ because
$\hat U_\text{rec}(t)$
 has a fixed non-zero error at fixed $t$. This
non-zero error may also become larger than the signal amplitude
$\| \hat H t \|$, explaining the increasing error as $t'-t \to 0$ for
some of the Hamiltonians.

Note that, from $\hat U(t'-t)$, we can also reconstruct Hamiltonians
that are time-dependent for times before $t$ and nearly constant
between $t$ and $t'$. In this way, stroboscopic reconstructions of a
time-dependent Hamiltonian may be obtained after large propagation
times.  Furthermore, $t/t_{n}<\pi$ becomes more restrictive as $n$
increases, thus the usefulness of taking measurements at two times
increases for larger systems.

\subsubsection{Enforcing a local reconstruction}

Of course, making use of additional information can only improve
the scheme. As an example, suppose that we know that the
Hamiltonian is nearest-neighbour only. One may then project
the reconstructed $\hat{H}_{\text{rec}}$ onto a nearest-neighbour
Hamiltonian. As can be seen in Fig.~\ref{fig:hamiltonian-reconstruction-twotimes},
this reduces the error dramatically.

\section{Conclusion and Outlook}

We studied in detail the application of recent scalable state
tomography models to quantum process tomography. At the hand of
unitary channels---quantum circuits such as logarithmic-depth
circuits, the quantum Fourier transform and unitary time-evolution
governed by one-dimensional local Hamiltonians---favourable scaling
with the number of qubits was numerically demonstrated. The scheme, as
presented, relies on an ancilla system, the preparation of a maximally
entangled state on the combined system, and local measurements after
application of the channel.  We also discussed an alternative scalable
scheme without the need for an ancilla, which displays the same
scaling properties as the ancilla-assisted scheme. The quality of the
reconstructed unitary channel may be quantified using the certificate
introduced in Ref.\ \cite{Cramer2010}.

We have also shown how one-dimensional local Hamiltonians may be
reconstructed from their corresponding unitary after an evolution time
inversely proportional in the size $n$ of the system and requiring
$\propto n^{2}$ measurements of each of $\propto n$ observables.  We
have discussed and numerically demonstrated how the restriction of
these small evolution times may be relaxed by taking measurements at
two different times. This enables the reconstruction and verification
of a quantum device at arbitrarily large times for as long as the
conditions for efficient state tomography are met, even if,
intermittently, the device has passed through highly entangled states.
Furthermore, the knowledge of the Hamiltonian being local may be
incorporated and has, for the Hamiltonians that we studied, improved
fidelities considerably.

Using the mixed-state tomography methods introduced in Refs.\
\cite{Baumgratz2013, Baumgratz2013a}, we expect non-unitary channels
to be similarly amenable to the scheme studied here.  Assessing the
quality of such a reconstruction, however, will rely on the ability to
quantify the quality of mixed-state reconstructions -- a goal that is,
in particular for many qubits and sufficient generality, still to be
met.

\begin{acknowledgments}
  This work was supported by an Alexander von Humboldt Professorship,
  the EU Integrating Project SIQS, the EU STREP EQUAM, the US-Army
  Research Office grant no. W91-1NF-14-1-0133, and the EPSRC
  (EP/K04057X/1). Numerical computations were performed on the
  bwUniCluster funded by the Ministerium f\"ur Wissenschaft, Forschung
  und Kunst Baden-W\"urttemberg %
  and the Universities of the State of Baden-W\"{u}rttemberg, Germany,
  within the framework program bwHPC.
\end{acknowledgments}

%



\appendix
\widetext

\section{Technical Details on Ancilla-Assisted Tomography and the Reduction of Experimental Effort}
\label{appendix:equivalence_details}
Let 
\begin{equation}
\ket{\Phi} = \frac{1}{2^{n/2}}\sum_{i=1}^{2^{n}}\ket{i}\ket{i} 
\end{equation}
be the maximally entangled state on the combined system. Then
\begin{equation}
\hat{\varrho}_{\mathcal
  E} = (\id_A \otimes \mathcal E)(\ket{\Phi}\!\bra{\Phi})=\frac{1}{2^{n}}\sum_{i,j}|i\rangle\langle j|\otimes  \mathcal E(|i\rangle\langle j|),
\end{equation}
i.e.,
$\langle i|\langle j|\hat{\varrho}_{\mathcal E}|i^\prime\rangle
|j^\prime\rangle=\langle j|\mathcal{E}(|i\rangle\langle
i^\prime|)|j^\prime\rangle/2^{n}$
and $\hat{\varrho}_{\mathcal E}$ thus completely characterizes the
channel $\mathcal E$.
We now review a known scheme (see, e.g.,
\cite{Flammia2012,Bendersky2008}) for obtaining measurement data on
$\hat \varrho_{\mathcal E}$ from measurements that use preperation and
measurement on the system without the ancilla. In addition, we show
that the linearly many product observables required for scalable state
tomography schemes \cite{Cramer2010,Baumgratz2013,Baumgratz2013a} may
be obtained from linearly many simple product preperations and product
measurements. In the simplest case, the schemes require the
observables
\begin{equation}
  \hat{P}_{k;\alpha_1,\dots,\alpha_r}=
  \id_{1,\dots,k}\otimes \hat{\sigma}^{\alpha_1}_{k+1}\otimes\cdots\otimes \hat{\sigma}^{\alpha_r}_{k+r}\otimes\id_{k+r+1,\dots,2n}\;\;\text{with}\;\; \alpha_i\in\{x,y,z\}\;\;\text{and}\;\;k=0,\dots,2n-r,
\end{equation}
where $\hat{\sigma}_i^x$, $\hat{\sigma}_i^y$, $\hat{\sigma}_i^z$ are
the Pauli matrices on site $i$. Measuring these observables, we obtain
the expectation values of the eigenprojectors
\begin{align}
  \hat{\Pi}_{k;\alpha_1,\ldots,\alpha_{r}}^{s_{1},\ldots,s_{r}}
  &=
    \id_{1,\ldots,k} \otimes \hat{\Pi}_{k+1;\alpha_{1}}^{s_{1}}
    \otimes \ldots \otimes
    \hat{\Pi}_{k+r;\alpha_{r}}^{s_{r}}
    \otimes
    \id_{k+r+1,\ldots,2n}, 
    \;\;\text{where}\;\;
    \hat{\Pi}^{s_{i}}_{k+i;\alpha_{i}} 
    =
    \ket{\alpha_{i},s_{i}}\bra{\alpha_{i},s_{i}},
    \quad
    s_{i} \in \{-1,1\},
\end{align}
with Pauli matrix eigenstates defined by
$\hat{\sigma}^{\alpha_{i}} \ket{\alpha_{i},s_{i}} = s_{i}
\ket{\alpha_{i},s_{i}}$.

For given
$\hat{\Pi}_{k;\alpha_1,\ldots,\alpha_{r}}^{s_{1},\ldots,s_{r}}$, write
$\hat{\Pi}_{k;\alpha_1,\ldots,\alpha_{r}}^{s_{1},\ldots,s_{r}}=\hat{P}_A\otimes
\hat{P}_S$
with the direct product referring to ancilla ($A$) vs. system
($S$). One finds
\begin{equation}
\begin{split}
\langle \hat{P}_A\otimes \hat{P}_S\rangle_{\hat{\varrho}_{\mathcal E}}&=\text{tr}[\hat{\varrho}_{\mathcal E}(\hat{P}_A\otimes \hat{P}_S)]=\frac{1}{2^{n}}\sum_{i,j,k,l}\langle i|\langle j| 
\bigl[\bigl(|k\rangle\langle l|\hat{P}_A\bigr)\otimes  \bigl(\mathcal E(|k\rangle\langle l|)\hat{P}_S\bigr)\bigr]
|i\rangle |j\rangle\\
&=\frac{1}{2^{n}}\sum_{i,j,k,l}\langle i| 
\bigl(|k\rangle\langle l|\hat{P}_A\bigr)|i\rangle\langle j| \bigl(\mathcal E(|k\rangle\langle l|)\hat{P}_S\bigr)
 |j\rangle
 =\frac{1}{2^{n}}\sum_{i,j,l}\langle l|\hat{P}_A|i\rangle\langle j| \mathcal E(|i\rangle\langle l|)\hat{P}_S
 |j\rangle\\
 &=\frac{1}{2^{n}}\sum_{i,l}\langle l|\hat{P}_A|i\rangle \text{tr}[\mathcal E(|i\rangle\langle l|)\hat{P}_S]
 =\frac{1}{2^{n}} \text{tr}[\mathcal E(\hat{P}_A^t)\hat{P}_S],
\end{split}
\end{equation}
where
$\hat{P}_A^t=\sum_{i,l}\langle l|\hat{P}_A|i\rangle |i\rangle\langle
l|$
is the transpose of $\hat{P}_{A}$ in the basis in which $\ket{\Phi}$
is entangled. Note that $\hat{P}_{A}$ is, up to a prefactor, a mixed
product state. The expectation value
$\langle \hat{P}_{A} \otimes \hat{P}_{S}
\rangle_{\hat{\varrho}_{\mathcal E}}$
may thus be obtained by preparing the state $\hat{P}_{A}$, sending it
through the channel and measuring an appropriate product of Pauli
matrices on the output state.

\section{Quantum Fourier Transform Gate Properties \label{appendix:qft-properties}}

The quantum Fourier transform circuit is given by
(cf. Eq.~\eqref{eq:circuits} in the main text)
\begin{align*}
  \widehat{\text{QFT}} &= \prod_{k=1}^{n} \Biggl[ \Biggl( \prod_{j=1}^{n-k}
  \widehat{\text{CR}}_{k,k+j}(\pi/2^{j}) \Biggr) \hat{H}_{k} \Biggr]
\end{align*}
As one two-qubit gate acts on every pair of qubits, the depth (defined
in Sec.~\ref{sec:reconstruct-unitaries} of the main text) at the split
$i|i+1$ is given by $d_{i} = i(n-i)$, the maximal depth is
$d_{\text{max}} = \max_{i} d_{i} = \lfloor \frac{n^{2}}{2} \rfloor$
and there is an MPO representation with bond dimension
$D \le 4^{d_{\text{max}}} = 4^{\lfloor n^{2}/2 \rfloor}$, which is not
efficient. However, we can obtain a smaller bond dimension as follows:
The conditional rotation gates are given by
$\widehat{\operatorname{CR}}_{k,k+j}(\phi) = \ket{0}_{k}\bra{0}
\otimes \id_{k+j} + \ket{1}_{k}\bra{1} \otimes R_{k+j}(\phi)$
with $R(\phi) = \ket{0}\!\bra{0} + \e^{\ii\phi} \ket{1}\!\bra{1}$.
Observe that
\begin{align*}
  \prod_{j=1}^{n-k} \widehat{\text{CR}}_{k,k+j}(\pi/2^{j})
  &=
    \ket{0}_{k}\!\bra{0} \otimes \id_{k+1,\ldots,n}
    +
    \ket{1}_{k}\!\bra{1} \otimes
    R_{k+1}(\pi/2) \otimes \ldots \otimes R_{n}(\pi/2^{n-k}).
\end{align*}
The two summands on the right-hand side are tensor products and thus
have an MPO representation with bond dimension 1. Therefore, the
left-hand side has an MPO representation with bond dimension 2
\cite{Schollwoeck2011}. Concatenating the $n$ terms of this form, we
obtain an MPO representation of $\widehat{\operatorname{QFT}}$ with
bond dimension $2^{n}$, which still is not efficient.

As mentioned in the main text, an approximation of the quantum Fourier
transform can be simulated classically with polynomial resources
\cite{Yoran2007}. Using numerical MPO compression
techniques~\cite{Schollwoeck2011}, we obtain a feasible approximate
MPO representation of the quantum Fourier transform circuit on
$n \le 32$ qubits: At bond dimension 16, the error is bounded by
$[2(1-\sqrt F)]^{1/2} < 2 \times 10^{-5}$.

Because the quantum Fourier transform circuit contains many small
conditional rotations, one can try to approximate the circuit by a
circuit $\widehat{\operatorname{QFT}}_{c}$ obtained by dropping all
conditional rotations from Eq.~\eqref{eq:circuits} in the main text
with $j > c$ \cite{Coppersmith1994}.  The operator norm error
satisfies (see also \cite{Yoran2007})
\begin{align*}
  \|\widehat{\operatorname{QFT}}-\widehat{\operatorname{QFT}}_{c}\|
  &\le
    \sum_{k=1}^{n} \sum_{j=c+1}^{n-k} \| \id - \widehat{\operatorname{CR}}(\pi/2^{j}) \|
    \le
    \pi \sum_{k=1}^{n} \sum_{j=c+1}^{n-k} \frac1{2^{j}}
    \le
    n \pi \sum_{j=c+1}^{\infty} \frac1{2^{j}}
    =
    \frac{n \pi}{2^{c}}
\end{align*}
The operator norm error in turn upper bounds our error measure
$[2(1-\sqrt F)]^{1/2}$: For two unitaries $U$, $U'$ and with the
Frobenius norm $\| M \|_{F} = \tr(M^{\dagger} M)$,
\begin{align*}
  2(1-\sqrt F) = \min_{\alpha\in\mathbb{R}}
  \| \ket{\psi_{U}} - \e^{\ii\alpha} \ket{\psi_{U'}} \|_{2}^{2}
  \le
  \| \ket{\psi_{U}} - \ket{\psi_{U'}} \|_{2}^{2}
  =
  \frac{\| U - U' \|_{F}^{2}}{2^{n}}
  \le
  \| U - U' \|^{2}.
\end{align*}

\section{Hamiltonian Reconstruction for Ising and Random Hamiltonians}
\label{appendix:other hamiltonians}

\begin{figure*}[t]
  \centering
  \includegraphics[width=\linewidth]{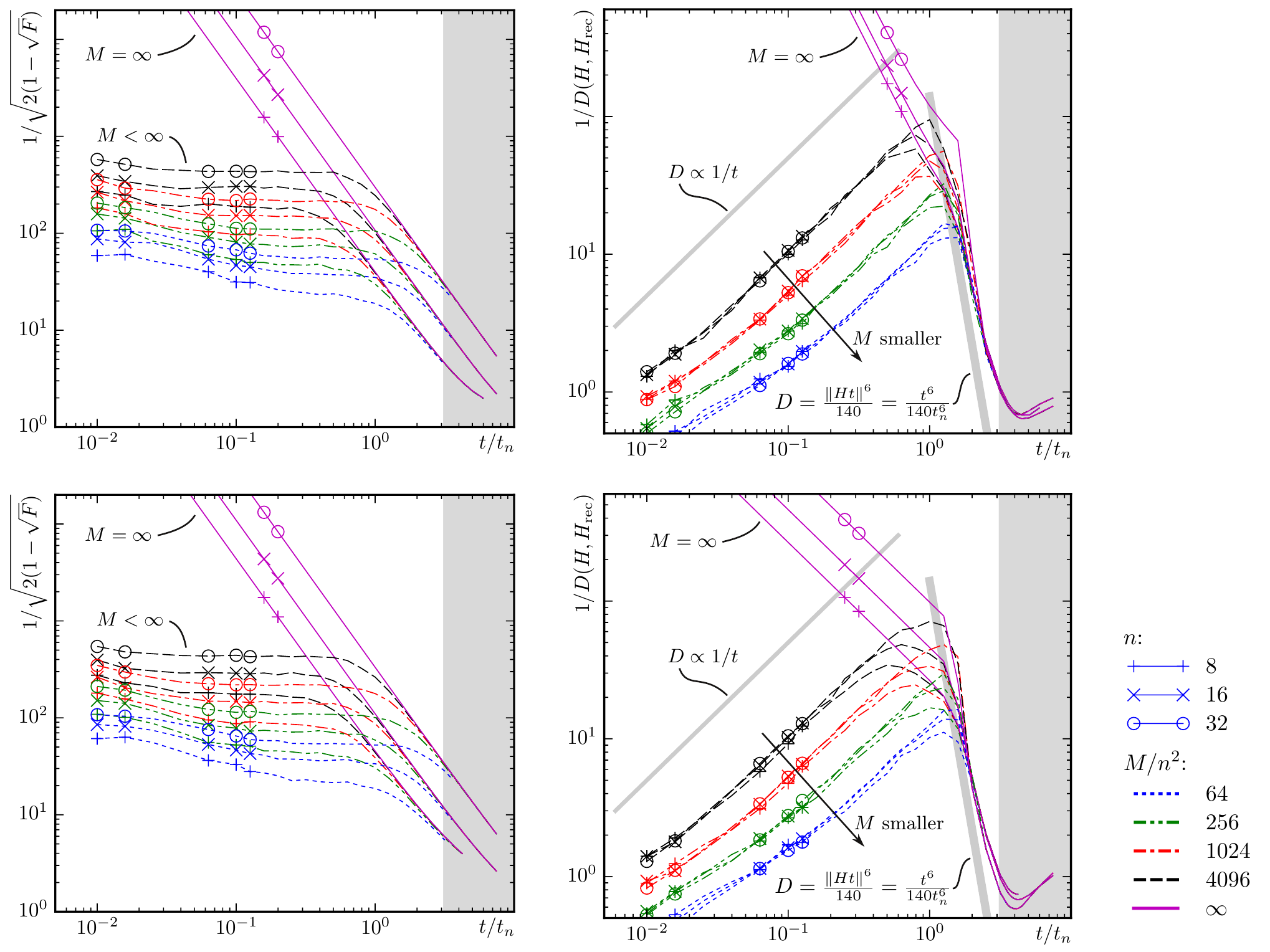}

  \caption{(Color online) \textit{Left:} Fidelity between exact state
    $\ket{\psi_{\mathcal E}}$ and tomographic estimate
    $\ket{\psi_{\mathcal E}^\text{rec}}$ for a unitary time evolution
    $\hat{U} = \e^{-\ii \hat{H} t}$ with $\hat{H}$ the Hamiltonian of
    the critical Ising model (top row) and a Hamiltonian with random
    nearest-neighbour interaction (bottom row). The Hamiltonians act
    on $n$ qubits and the tomographic estimate is based on complete
    measurements on blocks of $r = 3$ consecutive qubits with $M$
    measurements per observable.  \textit{Right:} Inverse relative
    distance between $\hat{H}$ and $\hat{H}_\text{rec}$ reconstructed
    from $\ket{\psi_{\mathcal E}^\text{rec}}$. The unit of time is
    $t_{n} = 1/\|\hat{H}\| \sim 1/n$.  Results are very similar to the
    isotropic Heisenberg data shown in
    Fig.~\ref{fig:hamiltonian-reconstruction} in the main text.}
  \label{fig:appendix-fig-other-hamiltonians}
\end{figure*}

Fig.~\ref{fig:hamiltonian-reconstruction} in the main text shows the
performance of our reconstruction scheme for local Hamiltonians using
the example of the isotropic Heisenberg
Hamiltonian. Fig.~\ref{fig:appendix-fig-other-hamiltonians} shows data
for the critical Ising model and for a Hamiltonian with random
nearest-neighbour interaction, with matrix elements chosen uniformly
from $[-1, 1]$. Reconstruction works equally well for the critical
Ising model and the randomly chosen nearest-neighbour interaction.

\begin{thebibliography}{53}
\bibitem [{ft:({\natexlab{b}})}]{ft:noancilla-statistical-errors}%
  \BibitemOpen
  \bibinfo {note} {{These statistical errors directly translate to statistical
  errors for the scheme without ancilla: With ancilla and for each observable
  $\hat{P}_{k;\alpha_{1},\ldots,\alpha_{r}}$, we simulate $M$ measurements.
  This translates to preparing each of the $2^{s}$ eigenstates of
  $\hat{\sigma}_{k-1}^{\beta_{1}} \otimes \cdots \otimes
  \hat{\sigma}_{k+s}^{\beta_{s}}$ an average of $M/2^{s}$ times and measuring
  corresponding local observables on $s'$ sites after application of the
  channel. Here, $s \le \lceil \frac r2 \rceil$ and $s+s' = r$.}}\BibitemShut
  {Stop}%
\bibitem [{ft:({\natexlab{f}})}]{ft:h-rec-scaling-prefactor}%
  \BibitemOpen
  \bibinfo {note} {{If the assumption is violated, we get an additional
  time-independent prefactor.}}\BibitemShut {Stop}%
\end{thebibliography}
\end{document}